\documentclass[12pt]{article}
\usepackage{hyperref}
\hypersetup{colorlinks,linkcolor={blue},citecolor={red},urlcolor={red}} 
\usepackage{scicite}
\usepackage{tikz,xcolor,hyperref}
\usepackage{graphicx}
\usepackage{amsmath}
\usepackage{amssymb}
\usepackage{float}
\usepackage[all]{xy}
\usepackage{amsfonts}
\usepackage{dcolumn}
\usepackage{bm}
\usepackage{xcolor}
\usepackage{caption}
\usepackage{hyperref}
\usepackage{multirow}
\usepackage[utf8x]{inputenc}
\usepackage{epstopdf}
\usepackage{capt-of}
\usepackage{dcolumn}   
\usepackage{bm}
\usepackage{dsfont}
\usepackage{relsize}
\usepackage{mathtools}
\usepackage{hyperref}
\hypersetup{colorlinks,linkcolor={blue},citecolor={red},urlcolor={red}} 
\usepackage{scicite}
\usepackage{tikz,xcolor,hyperref}

\definecolor{lime}{HTML}{A6CE39}
\DeclareRobustCommand{\orcidicon}{%
	\begin{tikzpicture}
	\draw[lime, fill=lime] (0,0) 
	circle [radius=0.20] 
	node[white] {{\fontfamily{qag}\selectfont \tiny ID}};
	\draw[white, fill=white] (-0.0625,0.095) 
	circle [radius=0.012];
	\end{tikzpicture}
	\hspace{-2mm}
}

\foreach \x in {A, ..., Z}{%
	\expandafter\xdef\csname orcid\x\endcsname{\noexpand\href{https://orcid.org/\csname orcidauthor\x\endcsname}{\noexpand\orcidicon}}
}

\usepackage{times}

\newenvironment{sciabstract}{%
\begin{quote} \bf}
{\end{quote}}

\newcounter{lastnote}

\title{Configurational entropy in $f(T)$ gravity} 

\author
{Snehasish Bhattacharjee \orcidA{}\\
\\
\normalsize{Department of Astronomy, Osmania University, Hyderabad, India, 500007}\\
\normalsize{snehasish.bhattacharjee.666@gmail.com}\\
\normalsize{}\\
\\
}

\date{\today}

\begin{document} 


\baselineskip24pt


\maketitle


\begin{sciabstract}
  The evolution of the configurational entropy of the universe relies on the growth rate of density fluctuations and on the Hubble parameter. In this work, I present the evolution of configurational entropy for the power law $f(T)$ gravity model of the form  $f(T) = \zeta (-T)^ b$, where, $\zeta = (6 H_{0}^{2})^{(1-s)}\frac{\Omega_{P_{0}}}{2 s -1}$ and $b$ a free parameter. From the analysis, I report that the configurational entropy in $f(T)$ gravity is negative and decreases with increasing scale factor and therefore consistent with an accelerating universe. The decrease in configurational entropy is the highest when $b$ vanishes since the effect of dark energy is maximum when $b=0$. Additionally, I find that as the parameter $b$ increases, the growth rate, growing mode and the matter density parameter evolve slowly whereas the Hubble parameter evolve rapidly. The rapid evolution of the Hubble parameter in conjunction with the growth rate for the $b=0$ may provide an explanation for the large dissipation of configurational entropy. 

\end{sciabstract}

\section{Introduction}
Predictions of $\Lambda$CDM cosmological model have been tremendously successful in describing the universe at all length scales \cite{ft1,ft2} by incorporating dark matter and dark energy \cite{ft3,ft4} where dark matter is responsible for providing the additional gravitational pull to keep the galaxies and clusters from flying apart while dark energy is responsible for fueling the acceleration of the universe at the largest scales. Nonetheless, after numerous attempts to detect these dark components, their identity remain elusive and their presence unconfirmed.
The other problem that gravely compromises the efficiency of the "standard" cosmological model is the so-termed Hubble tension, where the value of $H_{0}$ obtained from CMB anisotropies differ significantly from the one obtained from local observations \cite{ft6to9}. \\
In this spirit, extended theories of gravity have been formulated in which the gravitational sector of the field equations are altered keeping the matter-energy sector unchanged. In this work, I am going to work with teleparallel $f(T)$ gravity where $f(T)$ represents any arbitray function of torsion scalar $T$. Teleparallel gravity theories are formulated through the Weitzenb\"{o}ck connection \cite{ft18,ft19} which is curvature-less and satisfy the metricity condition \cite{jackson}. For $f(T)=T$, the dynamical equations produced for a Lagrangian consisting only of only $T$ are identical to the onces obtained from GR differing only by a boundary term $B$ and is termed Teleparallel equivalent of General Relativity (TEGR) \cite{jackson}. Teleparallel gravity have been reported to yield encouraging results in various cosmological scenarios \cite{ft18,ft32,ft33,sanjay,bhatta,baryo} in addition to galactic \cite{ft34} and solar system \cite{ft36} scales. In \cite{linder2}, the phenomena of late-time cosmic acceleration was addressed in the framework of $f(T)$ gravity.  In \cite{nunes}, the author studied the evolution of scalar perturbations in $f(T)$ gravity and its effects on the cosmic microwave background (CMB) anisotropy and concluded that a small deviation of the $f(T)$ gravity from the $\Lambda$CDM is preferred and also resolves the Hubble tension prevailing in the $\Lambda$CDM model. In \cite{bbnft} a complete analysis of power law, exponential and square-root exponential $f(T)$ gravity models were carried out considering both background and linear perturbation evolution and the free parameters appearing in the models were constrained at non-zero values in more than $3-\sigma$ confidence level, thus once again illustrating a preference of the observations for extended gravity. \\

In \cite{pandey}, the author conjectured that the evolution of the universe from a smooth to a highly non-linear state due to formation of structures is a direct consequence of the dissipation of configurational entropy of the universe. It has been argued that the present cosmic acceleration may be due to the depletion of configurational entropy \cite{framework}. A static universe with structures is unstable and results in a rapid depletion of configurational entropy. Thence, if the universe is assumed to be a thermodynamic system, the second law of thermodynamics must be obeyed and therefore it must expand to dampen out the formation of structures, since there exists no entropy generation processes which can counter the loss of configurational entropy \cite{framework}. The depletion of configurational entropy persists in a dust universe and damps out only in an accelerating universe \cite{framework}. \\

The manuscript is organized as follows: In Section \ref{sec2}, I present the theory of configurational entropy, growing mode and growth rate of density perturbations and $f(T)$ gravity. In Section \ref{sec3}, I present the results and in Section \ref{sec4} I summarize the results and conclude. Throughout the work I use $\Omega_{m0}=0.315$, $\Omega_{\Lambda}=1-\Omega_{m0}$ and $h=0.674$ \cite{planck}.

\section{Theory}\label{sec2}
\subsection{Evolution of configurational entropy} 
Let us consider a large comoving volume $V$ of the universe where  homogeneity and isotropy holds. The large volume is composed of smaller volume elements $dV$ with energy density $\rho (\overrightarrow{x},t)$ in each of them where $\overrightarrow{x}$ and $t$ denote the spatial coordinates and time assigned to each $dV$. \\
Bearing this in mind, the configurational entropy is defined as \cite{pandey} 
\begin{equation}
C (t) = - \int \rho \hspace{0.05cm}\text{log} \hspace{0.05cm} \rho d V .
\end{equation}
Note that the definition of configurational entropy is motivated from the definition of information entropy published by C. Shannon in 1948 \cite{shannon}. \\
The equation of continuity in an expanding universe reads 
\begin{equation}
\frac{\partial \rho}{\partial t} + \frac{1}{a} \bigtriangledown . (\rho \overrightarrow{\nu}) + 3 \frac{\dot{a}}{a} \rho = 0,
\end{equation}
where $\rho$ represent the density of the cosmological fluid, $a$ being the scale factor and $\overrightarrow{\nu}$ the peculiar velocity of each fluid element in $dV$. \\
Multiplying the above equation by $(1+ \text{log} \rho)$, integrating over $V$ and changing the variable from $t$ to $a$, yields \cite{pandey} 
\begin{equation}\label{1}
\frac{d C (a)}{d a} \dot{a} - G (a) + 3 \frac{\dot{a}}{a} C (a) = 0,
\end{equation}
where $G (a) = \frac{1}{a} \int \rho (\overrightarrow{x},a) \bigtriangledown . \overrightarrow{\nu} d V$, where $H = \frac{\dot{a}}{a}$ represents the Hubble parameter and $M = \int \rho (\overrightarrow{x},a) dV  =\int \overline{\rho} (1+\delta (\overrightarrow{x},a)) dV $ defines the total mass enclosed within the volume $V$ and $\partial  (\overrightarrow{x},a) = \frac{\rho ( (\overrightarrow{x},a) - \overline{\rho})}{\overline{\rho}}$ represents the density contrast with $\overline{\rho}$ being the mean density within the relevant comoving volume under inspection  \cite{framework}. \\
Under the framework of linear perturbation theory, the evolution of density fluctuations can be represented as $\delta (\overrightarrow{x},a) = D (a)\delta ( \overrightarrow{x})$, where $D(a)$ is defined as the growing mode of fluctuations. Keeping this in mind, the divergence of peculiar velocity $\nu$ can be defined as \cite{framework} 
\begin{equation}\label{2}
\bigtriangledown.\nu(\overrightarrow{x}) = -a \dot{a} \frac{d D (a)}{d a} \delta (\overrightarrow{x}).
\end{equation}
Substituting Eq. \ref{2} in Eq. \ref{1}, gives \cite{framework}
\begin{equation}
\frac{d C (a)}{d a} \dot{a} + \frac{1}{a} \overline{\rho} a \dot{a}    \frac{d D (a)}{d a} \left(D (a) \int\delta^{2}(\overrightarrow{x})dV + \int \delta(\overrightarrow{x}) dV  \right)  +3 \frac{\dot{a}}{a}( C (a) -M) = 0
\end{equation}
Since the second term inside the bracket vanishes upon integration, we finally end up with \cite{framework}
\begin{equation}\label{3}
\frac{d C (a)}{d a} + \overline{\rho} f    \frac{ D^{2} (a)}{ a} \int\delta^{2}(\overrightarrow{x})dV +  \frac{3}{a}( C (a)-M) = 0,
\end{equation}
where $f = \frac{d \text{ln}D}{d \text{ln}a}$ represents the dimensionless growth rate.\\
Solution of Eq. \ref{3} gives the evolution of configurational entropy in different cosmological models with respect to the standard $\Lambda$CDM cosmology. In this work, I numerically solve Eq. \ref{3} using $4^{th}$ order Runga-Kutta method to investigate the evolution of configurational entropy in a power-law $f(T)$ teleprallel gravity model and compare it with the same computed for the TEGR. Following \cite{framework}, I assume the time independent quantities in Eq. \ref{3} to be equal to 1 and set the initial scale factor $a_{i}= 10^{-3}$ with the initial condition $C(a_{i}) = M$. 
\subsection{Growing mode and growth rate of density perturbations}
Primordial density fluctuations grow with time due to gravitational instability causing the over-dense regions to become more over-dense and the under-dense regions to become more under-dense. The evolution of these perturbations can be studied with relative ease when $\delta \ll 1$. The growth equations for matter at sub-horizon scales reads \cite{linder/2003} 
\begin{equation}\label{4}
\delta^{''}(a) + \left[\frac{2-q}{a} \right] \delta^{'}(a) - \frac{3 \Omega_{m}}{2 a^{2}}\delta(a) =0,
\end{equation}
where primes denote derivatives with respect to the scale factor, $\Omega_{m}$ represents the matter density parameter and $q$ represents the deceleration parameter. \\
Assuming the matter content of the universe to be composed of usual baryonic and non-baryonic dark matter, the expression of the normalized Hubble parameter $E(a)=H(a)/H_{0}$ becomes
\begin{equation}\label{5}
E(a) = \frac{H(a)}{H_{0}} =  \sqrt{\Omega_{m} a^{-3} + (1-\Omega_{m})e^{3 \int_{a}^{1} [1+\omega(a^{'})]d \text{ln} a^{'}}},
\end{equation} 
where $\omega(a)$ is the EoS parameter which can be dynamical or a constant. In modified gravity, the effects of dark energy is purely geometric in nature.\\ Following \cite{linder/2003} I re-write Eq. \ref{4} as 
\begin{equation}\label{6}
D^{''} -\frac{3}{2}\left[\left(1-\frac{\omega(a)}{1+N(a)} \right) \frac{D^{'}(a)}{a}  + \frac{N(a)}{1+N(a)}\left(\frac{D}{a^{2}} \right) \right]=0, 
\end{equation}
where $D=\frac{\delta (a)}{\delta(a_{i})}$ and 
\begin{equation}
N(a) = \frac{\Omega_{m_{0}}(a)}{1- \Omega_{m_{0}}(a)}e^{-3 \int_{a}^{1} \omega(a^{'})d \text{ln} a^{'}}.
\end{equation}
From Eq. \ref{6} it is clear that the evolution of growth rate will be dictated by the parametrization of $\omega(a)$. In modified gravity, due to the presence of extra degrees of freedom, the growth rate will differ significantly if the free parameters are altered. In this work, I shall investigate the phenomena for a power-law $f(T)$ gravity. I numerically solve Eq. \ref{6} employing the $4^{th}$ order Runga-Kutta method and following \cite{framework} I normalize the solution such as $D(a_{0})=1$ in the case of TEGR. \\
The evolution of $ \Omega_{m}$ can be parametrized as 
\begin{equation}\label{7}
 \Omega_{m}(a) = \frac{1}{a^{3}}\left( \frac{ \Omega_{m_{0}}}{ E^{2}(a)}\right) 
\end{equation}
and the growth rate $f$ is defined as 
\begin{equation}\label{8}
f =  \Omega_{m}(a)^{\gamma},
\end{equation}
where $\gamma$ represents the growth index which in case of $\Lambda$CDM model is $\gamma\approx6/11$. However, $\gamma$ gets altered in the case of extended theories of gravity and usually incorporate the extra degrees of freedom appearing in the particular modified gravity model.  
\subsection{$f(T)$ teleparallel gravity}

The action in $f(T)$ gravity reads
\begin{equation}\label{9}
\mathcal{S} = \frac{1}{16 \pi G} \int d^{4}x e \left( T + \mathcal{L}_{m} + f(T) \right),
\end{equation}
where $e = \text{det}(e^{A}_{\mu})$ and $e_{A}(x^{\mu})$ are the vierbein fields and $\mathcal{L}_{m}$ represents matter Lagrangian. For a flat FRW spacetime with ($'+',\hspace{0.05in}'-',\hspace{0.05in}'-',\hspace{0.05in}'-'$) metric signature, the vierbein reads 
\begin{equation}\label{10}
e^{A}_{\mu} = \text{diag}(1,a,a,a).
\end{equation}
Upon varying the action (Eq. \ref{9}) with respect to the vierbeins, the field equations in this theory of gravity reads 
\begin{equation}\label{11}
4 \pi G e ^{\rho}_{A} \overset{em}{T}{}^{\nu}{}_{\rho} = e^{\rho}_{A}S_{\rho}^{\mu \nu} \partial_{\mu}(T) f_{TT} + e^{-1}\partial_{\mu}(ee^{\rho}_{A}S_{\rho}^{\mu\nu})[1+f(T)] - [1+f(T)]e^{\lambda}_{A}T^{\rho}{}_{\mu \lambda} S_{\rho}{}^{\nu\mu} + \frac{1}{4}e^{\nu}_{A}[T+f(T)],
\end{equation}
where $\overset{em}{T}{}^{\nu}{}_{\rho}$ represents the energy momentum tensor, $f_{T}= \partial f(T)/\partial T$ and $f_{TT}= \partial^{2} f(T)/\partial T^{2}$. Inserting Eq. \ref{10} in Eq. \ref{11}, the Friedmann equations read 
\begin{equation}\label{12}
H^{2} = \frac{1}{3}\left[8 \pi \rho G - T f_{T} - f \right] 
\end{equation}
and 
\begin{equation}\label{13}
\dot{H}=-\left[\frac{4 \pi G (p + \rho)}{ 2 T f_{TT} + f_{T}+1} \right].
\end{equation}
In FRW spacetime, $T=-6 H^{2}$ and therefore the normalized Hubble parameter can be re-expressed as 
\begin{equation}\label{14}
E(a)= \frac{H(a)}{H_{0}} = \sqrt{\frac{T(a)}{T_{0}}}.
\end{equation}
The effective EoS parameter ($\omega$) reads \cite{ft}
\begin{equation}\label{15}
\omega = -1-\frac{1}{3}\frac{d \text{ln}T}{d \text{ln}a}\frac{f_{T} + 2 T f_{TT}}{\left[ (f/T)-2f_{T}\right] }.
\end{equation}
Additionally, using the definition of $T$ along with Eq. \ref{14} gives \cite{ft}
\begin{equation}
\frac{d \text{ln}T}{d \text{ln}a} = 2 E (a)T_{0}\frac{d \text{ln}E}{d \text{ln}a}.
\end{equation}
The normalized Hubble parameter in $f(T)$ gravity can be expressed as \cite{ft}
\begin{equation}
E(a)\simeq\sqrt{\Omega_{m_{0}} a^{-3} + \Omega_{P_{0}}x(a)}
\end{equation}
where $\Omega_{P_{0}} = 1 - \Omega_{m_{0}}$ and
the function $x(a)$ is defined as \cite{ft}
\begin{equation}\label{16}
x(a)=\frac{(f - 2 T f_{T})}{\Omega_{P_{0}}T_{0}}.
\end{equation}

\subsubsection{Power-Law $f(T)$ model}
In the present work, I shall work with the power law model of Bengochea and Ferraro \cite{ferraro} defined as 
\begin{equation}\label{17}
f(T) = \zeta (-T)^ b
\end{equation}
where $b$ is the free parameter and 
\begin{equation}
\zeta = (6 H_{0}^{2})^{(1-s)} \frac{\Omega_{P_{0}}}{2 s -1}.
\end{equation}
Substituting this in Eq. \ref{15} and Eq. \ref{16}, I obtain 
\begin{equation}\label{18}
\omega = -1 + \frac{2b}{3}(1+z)\frac{d \text{ln}E}{d z}
\end{equation}
and 
\begin{equation}
x(a,b) = E^{2b}(a,b).
\end{equation}
Ref \cite{ft} derived the expressions for the normalized Hubble parameter $E(a,b)$ and the growth index $\gamma$ in power law model respectively as
\begin{equation}\label{19}
E (a,b) \simeq \sqrt{E^{2}(a) + \Omega_{P_{0}} \text{ln}\left[E^{2}(a) \right] b },
\end{equation} 
and
\begin{equation}\label{20}
\gamma \approx \frac{6}{11-6b}.
\end{equation}
Note that for $b=0$, $\gamma \approx 6/11$. 

\section{Results}\label{sec3}
Since the Hubble parameter is one of the most important parameter characterizing the growth of clustering, in the top left panel, I show the evolution of squared normalized Hubble parameter $E^{2}(a)$ (Eq. \ref{19}) for different values of $b$ where it is clearly observed that $E^{2}$ increases faster with decreasing scale factor as the free parameter $b$ increases from 0 towards unity. At $a=1$,  $E^{2}(a)$ for all $b$ as it should be. Note that the choice of the parameter $b$ is not arbitrary as a recent study constrained the power law $f(T)$ gravity model from big bang nucleosynthesis and imposed a upper limit of $b \lesssim 0.94$ \cite{sal}.\\
In the top right panel, I show the evolution of EoS parameter $\omega$ (Eq. \ref{18}) of the geometric dark energy arising from the power law model employed in the work where I find the EoS parameter to remain in quintessence region throughout the cosmic evolution. Additionally, the magnitude of $\omega$ increases with decreasing scale factor as the parameter $b$ increases. For $b=0.9$, $\omega$ approaches the dust case at early times.\\
In the middle left panel, I show the evolution of matter density parameter $\Omega_{m}(a)$ where it can be clearly noted that for the the TEGR case implied by $b=0$, the matter domination starts at an earlier epoch than when $b>0$. However, the profiles converge at $a=1$. \\
In the middle right panel, I show the evolution of squared growing mode $D^{2}(a)$ which I obtain after numerically solving Eq. \ref{6} using 4th order Runga Kutta method after the substitution Eq. \ref{18} in it. I also normalize $D^{2}(a_{0})=1$ for the TEGR case. It can be noted that the slope of the growing mode is the steepest for the TEGR ($b=0$) case and flattens out as the parameter $b$ increases indicating that the presence of power law $f(T)$ gravity diminishes the amplitude of the perturbations at all times with its effect increasing as the parameter $b$ increases. Additionally, at smaller values for the scale factor $a$, the profiles start to flatten out indicating the minute but non-negligible contribution of dark energy component on the formation of large cosmic structures.\\
Lower left panel shows the evolution of dimensionless growth rate $f(a)$ which I obtain using Eq. \ref{7} and Eq. \ref{20}. Such a parametrization of $f(a)$ employing the matter density history $\Omega_{m}$ and the growth index $\gamma$ allows it to encompass the information regarding the expansion and the growth history. In this work, the only free parameter is $b$ which upon altering significantly affects the evolution of $f(a)$ as depicted in the figure. The amplitude of the growth rate is significantly affected in power-law $f(T)$ gravity with the amplitude diminishing when the parameter $b$ increases. This indicates that for the TEGR case, the effect of dark energy is the highest and acts against gravity and therefore suppresses the growth of density perturbations. This effect however reduces once the parameter $b$ is allowed to be non zero. At very early times $i.e, a \ll 1$, $f(a)$ tends towards unity in all cases.\\
Lastly, in the lower right panel I show the evolution of configurational entropy $C(a)$ by solving Eq. \ref{3} and Eq. \ref{6} using $4^{th}$ order Runga-Kutta method. In this work, following the prescription of \cite{framework}, I set $C(a_{i})=M=2$. Note that the other two possible cases are $C(a_{i})<M$ and $C(a_{i})>M$. Interestingly, the configurational entropy in $f(T)$ gravity is negative and decreases with increasing scale factor. The decrease in configurational entropy is the highest for the TEGR case and can be attributed to the fact that the growth rate is the maximum for the TEGR case which may lead to a larger dissipation of $C(a)$. The dissipation of $C(a)$ is largely influenced by the free parameter $b$, where an increase in $b$ reduces the dissipation of $C(a)$.  

\begin{figure}[H]
\centering
  \includegraphics[width=7.9 cm]{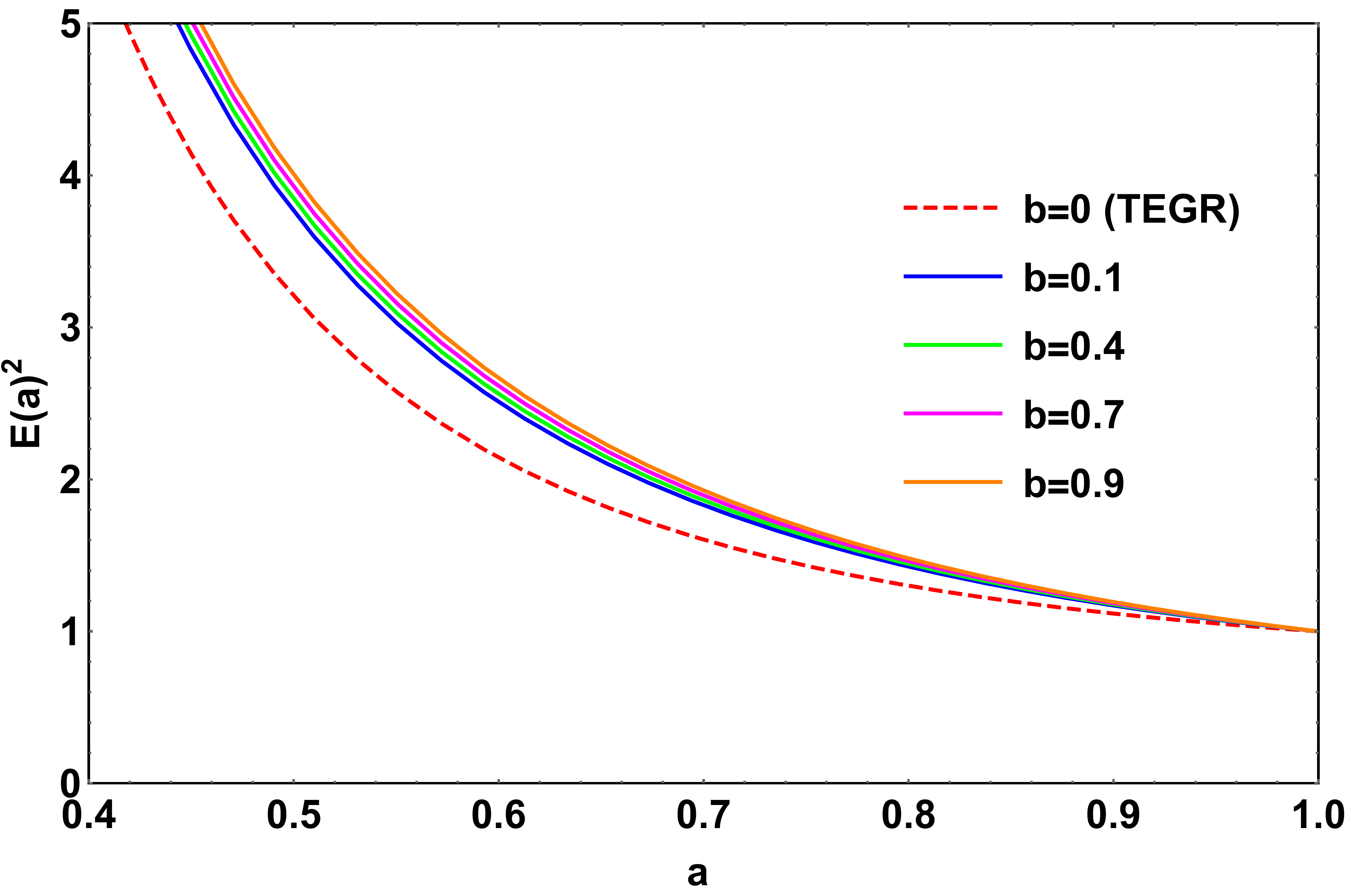}
  \includegraphics[width=7.9 cm]{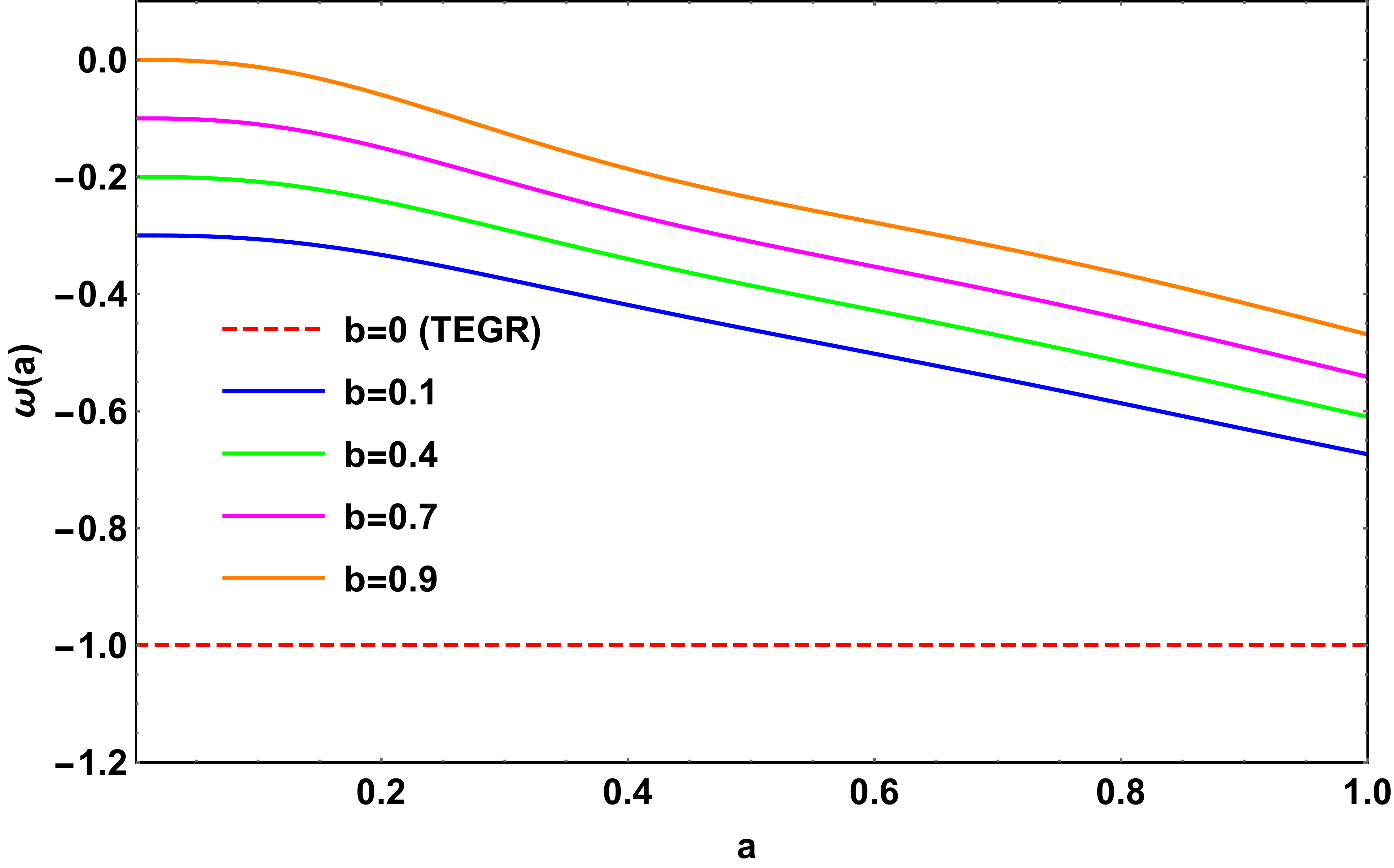}
  \includegraphics[width=7.9 cm]{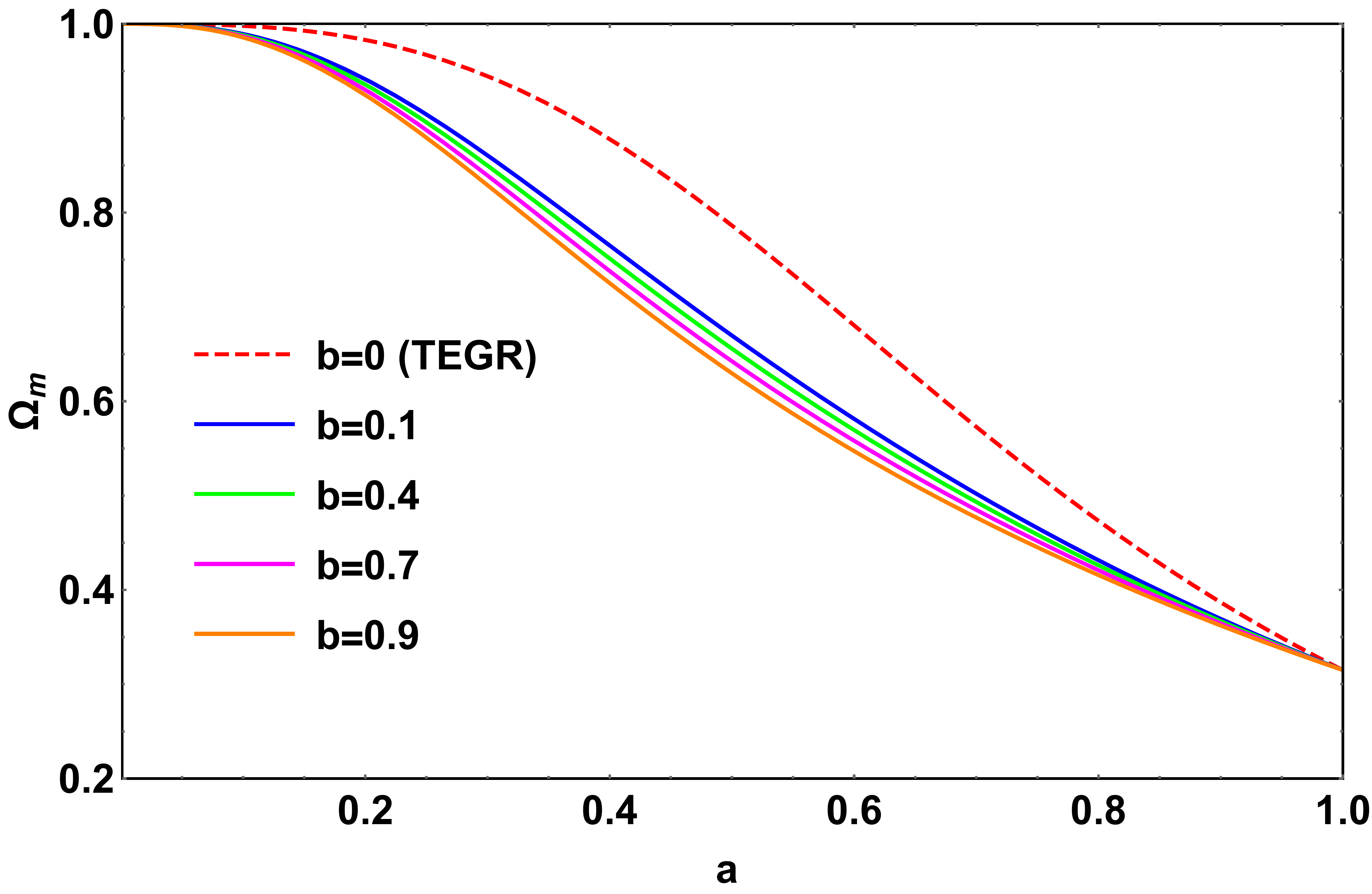}
  \includegraphics[width=7.9 cm]{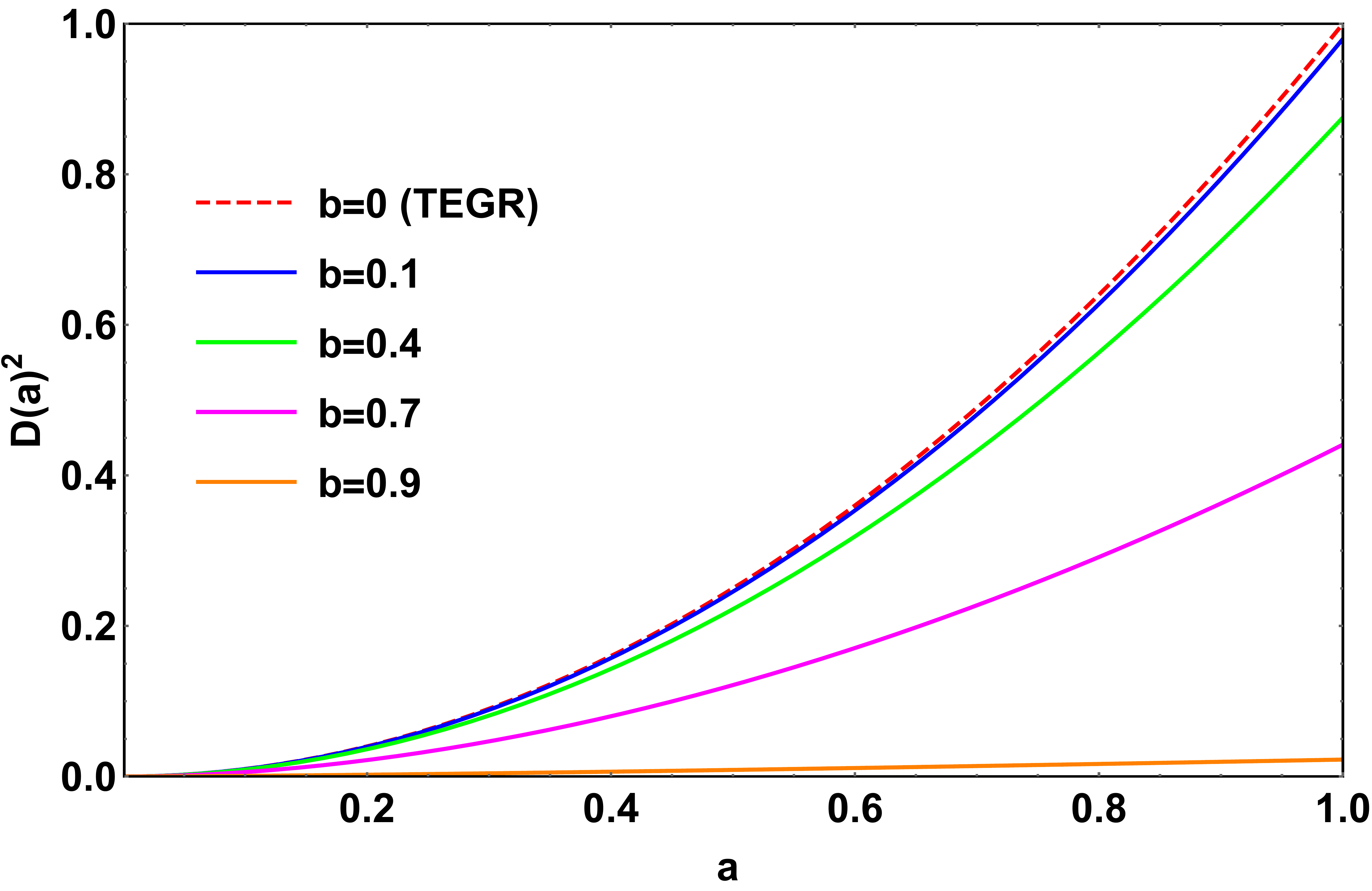}
  \includegraphics[width=7.9 cm]{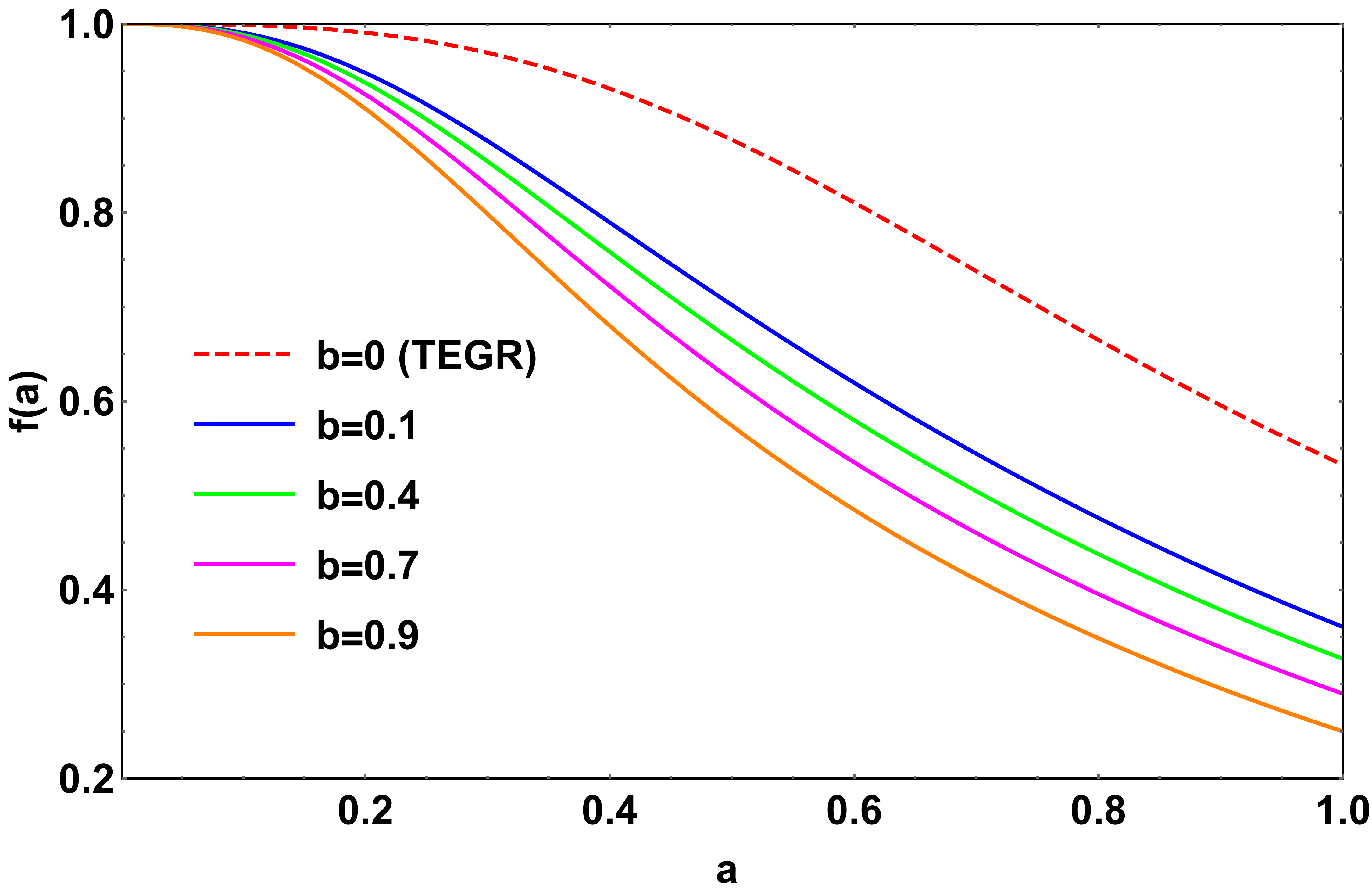}  
  \includegraphics[width=7.9 cm]{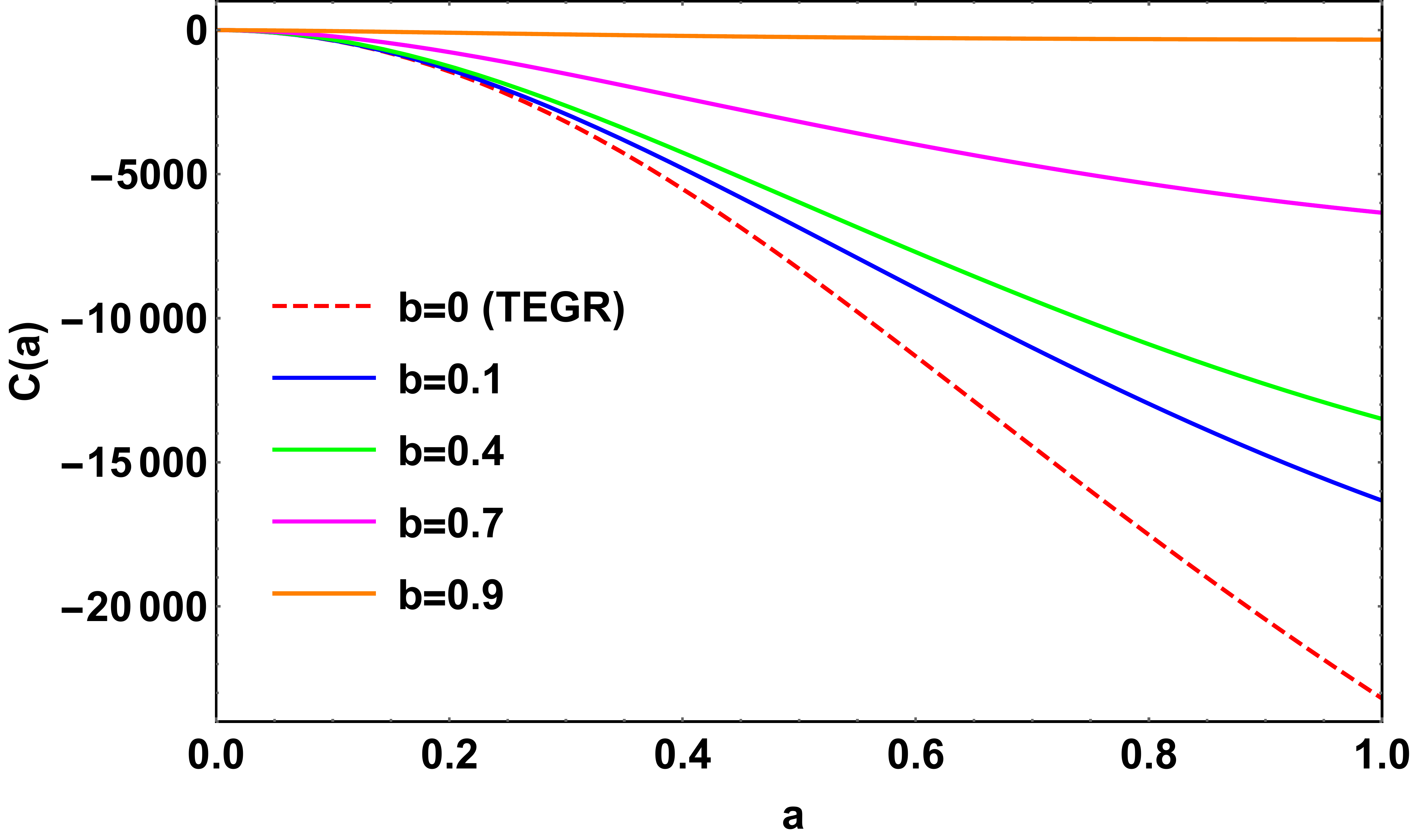}
  
\caption{Top left panel shows the evolution of squared normalized Hubble parameter $E^{2}(a)$, top right panel shows the evolution of dark energy EoS parameter $\omega(a)$, middle left panel shows the evolution of matter density parameter $\Omega_{m}(a)$, middle right panel shows the evolution of squared growing mode $D^{2}(a)$ where $D=\frac{\delta (a)}{\delta(a_{i})}$, lower left panel shows the evolution of growth rate $f(a)$ and lower right panel shows the evolution of configurational entropy $C(a)$ as a function of $a$ for different values of $b$.}
\label{FIG1}
\end{figure}

\section{Conclusions}\label{sec4}
Configurational entropy is an interesting concept in cosmology. The depletion of configurational entropy is directly linked to an universe being evolved from a highly smooth state to a non-linear clumpy state observed at the present epoch due to the formation of large structures. In \cite{framework}, the authors argued that the cosmic acceleration could be a reminiscent of large depletion of configurational entropy. \\
In this paper, using the framework of \cite{pandey,framework}, I investigated the evolution of Hubble parameter, matter density parameter, EoS parameter, linear growth rate, density perturbations and most importantly the configurational entropy for the power law model $f(T) = \zeta (-T)^ b$, where, $\zeta = (6 H_{0}^{2})^{(1-s)} \frac{\Omega_{P_{0}}}{2 s -1}$ and $b$ a free parameter. The results can be summarized as follows: \\
The profiles for the squared normalized Hubble parameter $E^{2}(a)$ is sensitive to the power law index $b$ and it is found that $E^{2}(a)$ increases faster with decreasing scale factor as $b$ increases. \\ The EoS parameter $\omega$ of the geometric dark energy arising due to power-law modifications of TEGR never cross the phantom divide line and remain in quintessence region throughout the cosmic evolution. Additionally, the profiles approach toward a dust universe faster as $b$ increases. \\ The matter density parameter $\Omega_{m}(a)$ for power law model decreases faster with increasing scale factor for the TEGR case and this can be purely attributed to the fact that for the TEGR case the effect of the dark energy is maximum as evident form the plot of $\omega(a)$. \\     The slope of the growing mode is the steepest for the TEGR ($b=0$) case and flattens out as the parameter $b$ increases indicating that the presence of power law $f(T)$ gravity diminishes the amplitude of the perturbations at all times with its effect increasing as the parameter $b$ increases. Additionally, at smaller values for the scale factor $a$, the profiles start to flatten out indicating the minute but non-negligible contribution of dark energy component on the formation of large cosmic structures.\\
The evolution of growth rate is significantly affected in the presence of power-law $f(T)$ gravity with the amplitude of $f(a)$ diminishing as the parameter $b$ increases. This once again highlights and reassures the fact for the TEGR case, the effect of dark energy is the severest and therefore contributes tremendously in suppressing the growth of density perturbations. This effect however reduces once the parameter $b$ is allowed to be non zero. At very early times $i.e, a \ll 1$, $f(a)$ tends towards unity in all cases.\\
The configurational entropy in $f(T)$ gravity is negative and decreases with increasing scale factor and therefore consistent with an accelerating universe. The decrease in configurational entropy is the highest for the TEGR case since the effect of dark energy is maximum when $b=0$ and therefore can also be attributed to the fact that the growth rate is the maximum for the TEGR case which may lead to a larger dissipation of $C(a)$. The dissipation of $C(a)$ is largely influenced by the free parameter $b$, where an increase in $b$ reduces the dissipation of $C(a)$.  \\
In \cite{linder3}, a thorough investigation of exponential gravity characterized by an exponential in the Ricci scalar $R$ to address the cosmic acceleration was reported. In a future work, I shall try to analyze the evolution of configurational entropy for such models which most certainly would yield engrossing results and discussions.
\section*{Acknowledgments}
I thank the anonymous referee for useful criticisms and encouraging comments that helped me to improve the work significantly.

\end{document}